\documentclass[twoside,leqno,twocolumn]{article} 
 
\usepackage{ltexpprt} 
\usepackage{amsmath,amsfonts,amssymb}

\usepackage{color}
\usepackage[toc,page]{appendix}

\usepackage[notheorems]{dgleich-math}

\usepackage{graphicx,epsfig}
\usepackage{graphics,dcolumn,bm,epic,eepic,float}
\usepackage{amssymb,amsmath,multirow,rotate,color}
\usepackage{subfigure}
\usepackage{epstopdf}
\usepackage{tabularx}
\usepackage{booktabs}
\usepackage{paralist}

\usepackage{url}
\usepackage{amssymb}
\usepackage{graphicx}
\usepackage{helvet}
\usepackage{courier}
\usepackage{subfigure}
\usepackage{amsmath}
\usepackage{multirow}
\usepackage{comment}
\usepackage{amssymb}
\usepackage{multirow}
\usepackage{color}
\usepackage[ruled,linesnumbered,noend]{algorithm2e}%%
\usepackage{graphics}
\usepackage{rotating}
\usepackage{wrapfig}
\usepackage{hyperref}
\usepackage{xspace}
\hypersetup{colorlinks=true,urlcolor=blue,citecolor=blue,linkcolor=blue}

\usepackage{amsmath}
\usepackage{amssymb}
\usepackage{textcase}
\usepackage{soul}
\usepackage{dashrule}

\usepackage{url}

\newcommand{\be}{\begin{equation}}
\newcommand{\ee}{\end{equation}}
\newcommand{\bea}{\begin{eqnarray}}
\newcommand{\eea}{\end{eqnarray}}
\newcommand{\bit}{\begin{itemize}}
\newcommand{\eit}{\end{itemize}}

\newcommand{\cm}[1]{}
\newcommand{\jt}[1]{}
\usepackage{microtype}
\usepackage{relsize}
\usepackage[table]{xcolor}
\definecolor{lightblue}{rgb}{0.93,0.95,1.0}
\definecolor{lightgray}{rgb}{0.93,0.93,0.93}

\newcommand\TT{\rule{0pt}{2.3ex}}

\begin{document}

\title{What if \textsc{clique} were fast? Maximum Cliques in Information Networks and Strong Components in Temporal Networks}

\author{Ryan A. Rossi, David F.~Gleich, Assefaw H.~Gebremedhin \thanks{Purdue University}\\
\and
Md.~Mostofa Ali Patwary \thanks{Northwestern University}}
\date{}

\maketitle

\begin{abstract} \small\baselineskip=10pt 
Exact maximum clique finders have progressed to the point where we can
investigate cliques in million-node social and information networks, as well as 
find strongly connected components in temporal networks.
We use one such finder to study a large collection of modern networks emanating from 
biological, social, and technological domains.
We show inter-relationships between maximum cliques and several other 
common network properties, including network density, maximum core, and number of triangles.   In temporal networks, we find that the largest temporal strong components have around 20-30\% of the vertices of the entire network.  These components represent groups of highly communicative individuals.
In addition, we discuss and improve the performance and utility of the maximum clique finder itself.

\end{abstract}


\section{Introduction}
The \textsc{max-clique} problem is simply stated: 
\begin{quote}
 Consider a graph $G$.  A clique of size $k$ in $G$
is a subset of $k$ vertices that forms a complete
subgraph. The maximum clique problem is to find
the largest $k$ such that there is a clique of size $k$ in $G$.
\end{quote}
Determining if a graph has a clique of size $k$ is
well known to be NP-complete.  Approximating the largest clique is
also known to be hard~\cite{Khot-2001-cliques}.  Clique is even hard
for networks with power-law degree distributions and large clustering
coefficients~\cite{Abello-2006-clique-is-still-hard}.
Clearly, \textsc{max-clique} is not, generally, fast.  Yet, many real-world problems
have features that do not elicit worst-cast behaviors from well-designed algorithms.  
In this manuscript, we use a state-of-the-art \emph{exact maximum clique finder}~\cite{patt2012cliques} to investigate cliques in social and information networks.\footnote{We use the terms {\em graph} and {\em network} interchangeably throughout the manuscript.  Also, \emph{vertex} and \emph{node} are interchangeable too.}
We also use a relationship between strongly connected components in temporal networks and maximum cliques in a reachability graph to find these temporal strong components \cite{bhadra2003complexity,nicosia:023101}.

We have three goals with this manuscript.  

The first is to learn about the structure of social and information networks by comparing 
maximum clique size to other common network properties. 
We use 38 networks for this study, obtained from various domains,
that vary in size from a few hundred edges to 100 million edges.  
The results are presented in Section~\ref{sec:maxcliques-real}.  
We find interesting relationships between the average degree and maximum clique for network of Facebook friendships. Also, we study various upper-bounds on the maximum clique size for these networks.
Given a connected graph $G$, it can be shown that the {\em clique number} 
$\omega(G)$---the size of the largest clique in $G$---obeys the following chain of bounds
\[\delta(G)  \leq \omega(G) -1  \leq K(G)  \leq \Delta(G) \]
where $K(G)$ is the largest $k$ for which there is a $k$-core in $G$ 
(a $k$-core in $G$ is a maximal induced subgraph of $G$ in which every node has at least 
$k$ neighbors),  $\delta(G)$ is the minimum degree in $G$, and $\Delta(G)$ is the maximum degree in $G$.
Further, for a connected graph containing at least one triangle,
we can also show that \( \omega(G) \leq  \sqrt{2\cdot T(G)}\),
where $T(G)$ is the maximum number of triangles incident on any node in $G$.
We  find that, for most real-world networks,
$K(G)$ is typically within a factor of five of $\omega(G)$ whereas the ratio
between $\sqrt{2\cdot T(G)}$ and $\omega(G)$ can be fairly large. 

Inspired by \cite{nicosia:023101,bhadra2003complexity}, 
our second goal is to compute the largest temporal strong components.  When each edge represents a contact---a phone call, an email, or physical proximity---between two entities at a specific point in time, we have a particular type of evolving network structure~\cite{ferreira2002models}.  
A temporal path in such a network represents a sequence of contacts that obeys time.  A temporal strong component is a set of vertices where all temporal paths exist.  See Section~\ref{sec:tscc} for a formal treatment.  
Just like \textsc{clique}, checking if an evolving network has a temporal strong component of size $k$ is NP-complete.  Yet, temporal strong components correspond to cliques in a temporal reachability graph where each edge signifies the existence of a temporal path between vertices.  Thus, finding maximum cliques enables us to find temporal strong components.  We investigate the size and structure of these cliques on phone call networks and Twitter retweet networks in Section~\ref{sec:tscc}.  
Perhaps surprisingly, the largest temporal strong components typically involve 
only a small fraction (20\% to 30\%) of the total vertices.

For a small portion of these experiments, where the networks are relatively dense, we find that the exact maximum clique method requires unacceptable time.  
In Section~\ref{sec:clique-algs}, we discuss extensions
we made to the algorithm from~\cite{patt2012cliques} to reduce its runtime and 
increase its flexibility; in particular, we parallelize the clique-searches and use better upper-and-lower bounds on the maximum clique size to prune the search space further and allow additional flexibility. 

We make all the software underlying this manuscript available in the spirit of reproducible research.  See:
\centerline{\footnotesize
\url{http://www.cs.purdue.edu/~dgleich/codes/cliques}}

\section{Maximum Cliques in Social and Information Networks} 

\label{sec:maxcliques-real}

We begin by studying the properties of maximum cliques on modern networks.  
See Section~\ref{sec:clique-algs} for a discussion of the maximum clique procedure we utilize in this section.  

\subsection{Data.}
For all of these experiments, we discard any edge weights, self-loops, and only consider the largest strongly connected component.  In contrast to the temporal components we describe in the next section, in this section we mean the standard strong components.  If the graph is directed, we then remove non-reciprocated edges.
We present a complete summary of some basic network statistics for the following datasets in the supplementary material.  In total, we consider 8 broad classes.

\textbf{1.~Biological networks.}
We study a network where the nodes are proteins and the edges represent
protein-protein interactions (yeast~\cite{jeong2001lethality}).
We also study the celegans metabolic network~\cite{duch2005community} where the nodes are substrates and the edges are metabolic reactions.  Cliques in these networks represent biologically relevant modules.

\textbf{2.~Interaction networks.}
Here, nodes represent individuals and edges represent interaction in the form of
(i) email exchanges (email-enron~\cite{leskovec2009community}, email-enron-employ\cite{cohen2005enron}),
(ii) message posts (wiki-Talk~\cite{leskovec2010predicting}, fb-messages~\cite{opsahl2009clustering}),
(iii) cellphone calls (reality~\cite{eagle2006reality}), or
(iv) face-to-face contacts (infect-dublin, infect-hyper~\cite{isella2011s}).

\textbf{3.~Social networks.}
Nodes are again people, and edges represent social relationships in terms of
(i) importance (wiki-Vote~\cite{leskovec2010predicting}),
(ii) trust (epinions \cite{richardson2003trust}), 
(iii) friendship or follower (orkut \cite{mislove2007measurement}, 
  youtube \cite{mislove-2009-socialnetworksthesis}, 
  soc-LiveJournal~\cite{backstrom2006group},  
  flickr~\cite{gleich-flickr12},
  brightkite~\cite{cho2011friendship}, 
  gowalla~\cite{cho2011friendship}, 
  slashdot \cite{leskovec2009community}).

\textbf{4.~Facebook networks.}
The nodes are people and edges represent  ``Facebook friendships'' 
(fb-Stanford, fb-Penn94, fb-Berkeley, fb-Duke, fb-Texas84 \cite{traud2011social}). %Slashdot also has foe edges.

\textbf{5.~Technological networks.}
The nodes in these networks are either 
agents (p2p-Gnutella \cite{matei2002mapping}) or
routers (Internet-AS \cite{routeviews}, as-Skitter \cite{skitter}, WHOIS \cite{whois}, routers-rf\cite{rocketfuel}),
and edges are observed communications between the entities.

\textbf{6.~Collaboration networks.}
These are networks in which nodes represent individuals
and edges represent scientific collaborations
(MathSciNet~\cite{palla2008fundamental}, ca-CondMat, ca-AstroPh, ca-HepPh~\cite{leskovec2007graph}).
Large cliques in these networks are expected because they are formed when papers have many authors.  Thus, cliques are only interesting if they are larger than the author list of any particular paper.

\textbf{7.~Web link networks.}
Here, nodes  are web-pages and edges are hyperlinks between pages 
(polblogs~\cite{adamic2005political}, wikipedia~\cite{gleich2010tracking}, web-Google~\cite{leskovec2009community}). The largest clique represents the largest set of pages where the author wishes users to freely move around.

\textbf{8.~Retweet networks.}
Finally, in retweet networks, nodes are Twitter users and edges represent whether the users have retweeted each other (political-retweets~\cite{conover2011political}, twitter-copen~\cite{ahmed2010time}).  We generated twitter-crawl ourselves, see the temporal network datasets for more information.  Cliques are groups of users that have all mutually retweeted each other, and may represent an interest cartel.

\subsection{Analysis.}

The time taken for the exact clique finder, a clique heuristic, the size of the maximum clique, and the size of the heuristic clique are shown in Table~\ref{table:mc_data}.  
The identified cliques range in size from 3 nodes to 239 nodes.  In all but one case the clique finder from~\cite{patt2012cliques} identified the largest clique.  Even when it failed to find the largest clique on the flickr network, we have an upper-bound from the max $k$-core that the clique is no larger than 309 vertices.  In general, the method worked quickly, however, on the 106M edge orkut social network, it took almost 10 hours to find it.  

There are some interesting surprises in these clique numbers.  First, the technological networks have large cliques with 50 and 60 vertices. 
We did not expect to find such large cliques in these networks as cliques ought to represent redundant edges that could be more efficiently handled by alternative topologies.  Note that this finding may represent a by-product of the data collection methodology, rather than a property of the network.  Second, the largest clique in a social network or Facebook network has 51 members.  These are fairly large groups for everyone to know each other.  Third, and finally, the collaboration networks had large cliques, as we expect.  We suspect that the clique with 239 members represents a single paper.

\begin{table}[t!]
\caption{\textbf{Max Cliques in Real-world Networks}. }
\vspace{1mm}
\label{table:mc_data}
\centering\small
\scriptsize
%\begin{tabular}{l !{\vrule width 0.1mm} cc!{\vrule width 0.3mm}rr}
\begin{tabular}{@{}lrr cc rr@{}}
\TT  & \multicolumn{2}{c}{} &\multicolumn{2}{c}{ \textbf{Max Clique}} & \multicolumn{2}{c}{\textbf{Time} (seconds)}\\ 
\cline{2-7}
%\TT \textbf{Graph} &  \textsf{\textbf{Exact}}  & \textsf{\textbf{Heur}} & \textsf{\textbf{Exact}} & \textsf{\textbf{Heur}} \\ 
\TT \textbf{Graph} & $|V|$ & $|E|$ & \textsf{\textbf{Exact}}  & \textsf{\textbf{Heur}} & \textsf{\textbf{Exact}} & \textsf{\textbf{Heur}} \\ 
\noalign{\hrule height 0.3mm}
\TT
   \textsc{yeast} & 1458 & 1948 & 6 & 6 & 0.04  & 0.05\\ 
   \textsc{celegans} &  453 & 2025 & 9 & 8 & 0.04  & 0.04\\ 
\cline{1-7}
\TT 
   \textsc{mathscinet} & 332K & 820K & 25 & 25 & 0.05   & 0.07\\ 
   \textsc{ca-condmat} & 21K & 91K & 26 & 26 & 0.03   & 0.02\\ 
   \textsc{ca-astroph} & 17K & 196K & 57 & 57 & 0.08  & 0.04\\ 
   \textsc{ca-hepph} & 11K & 117K & 239 & 239 & 0.11   & 0.21\\ 
\cline{1-7}
\TT
   \textsc{fb-messages} & 1266 & 6451 & 5 & 5 & 0.03   & 0.01 \\
   \textsc{reality} & 6809 & 7680 & 5 & 5 & 0.05  & 0.04\\ 
   \textsc{enron-only} & 143 & 623 & 8 & 8 & 0.06  & 0.04\\ 
   \textsc{infect-hyp} & 113 & 2196 & 15 & 13 & 0.03   & 0.04\\
   \textsc{wiki-talk} & 92K & 360K & 15 & 12 & 1.52  & 0.06\\ 
   \textsc{infect-dub} & 410 & 2765 & 16 & 16 & 0.04   & 0.04\\ 
   \textsc{enron-large} & 33K & 180K & 20 & 18 & 0.41   & 0.06\\    
%\cline{1-7}
%\TT
%   \textsc{amazon} & 91K & 125K & 5 & 5 & 0.04   & 0.05\\ 
%\cline{1-7}
%\TT
%   \textsc{roadnet-pa} & 1M & 1.5M & 4 & 4 & 0.24   & 0.34\\ 
%   \textsc{roadnet-ca} & 1.9M & 2.7M & 4 & 4 & 0.51   & 0.6\\ 
\cline{1-7}
\TT
   \textsc{retweet-pol} & 96 & 117 & 4 & 3 & 0.03   & 0.02\\ 
   \textsc{twitter-cop} & 761 & 1029 & 4 & 4 & 0.06  & 0.06 \\ 
   \textsc{retweet-cra} & 1.1M & 2.2M & 13 & 13 & 0.74   & 0.22 \\ 
\cline{1-7}
\TT
   \textsc{wiki-vote} & 889 & 2914 & 7 & 6 & 0.03  & 0.06\\ 
   \textsc{epinions} & 26K & 100K & 16 & 12 & 0.08   & 0.04 \\ 
   \textsc{youtube} & 495K & 1.9M & 16 & 13 & 3.12   & 0.42 \\ 
   \textsc{slashdot} & 70K & 358K & 26 & 23 & 8.54   & 0.07 \\ 
   \textsc{brightkite} & 56K & 212K & 37 & 33 & 5.48   & 0.06 \\ 
   \textsc{flickr} & 513K & 3.1M & - & 35 & -   & 0.83 \\ 
   \textsc{gowalla} & 196K & 950K & 29 & 29 & 0.68   & 0.13 \\ 
   \textsc{orkut} & 2.9M & 106M & 47 & 43 & 32238.8   & 45.34 \\ 
   \textsc{livejournal} & 4M & 27.9M & 214 & 214 & 2.47   & 2.47 \\ 
\cline{1-7}
\TT
   \textsc{duke} & 9885 & 506K & 34 & 29 & 153.81   & 0.06 \\ 
   \textsc{berkeley} & 22K & 852K & 42 & 36 & 32.19   & 0.17 \\ 
   \textsc{penn} & 41K & 1.3M & 44 & 42 & 21.25   & 0.22 \\ 
   \textsc{stanford} & 11K & 568K & 51 & 46 & 655.57   & 0.11 \\ 
   \textsc{texas} & 36K & 1.5M & 51 & 43 & 3543.86  & 0.24 \\ 
\cline{1-7}
\TT
   \textsc{p2p-gnutel} & 62K & 147K & 4 & 4 & 0.04   & 0.06 \\ 
%   \textsc{iptrace-B} & 143K & 203K & 5 & 5 & 0.13   & 0.05 \\ 
%   \textsc{iptrace} & 1M & 1.9M & 8 & 6 & 55.58   & 0.5 \\ 
   \textsc{internet-as} & 40K & 85K & 16 & 11 & 0.16   & 0.02 \\ 
   \textsc{routers-rf} & 2113 & 6632 & 16 & 15 & 0.07   & 0.05 \\ 
   \textsc{whois} & 7476 & 56K & 58 & 51 & 9708.89   & 0.05 \\ 
   \textsc{as-skitter} & 1.6M & 11M & 67 & 61 & 1784.65   & 2.33 \\ 
\cline{1-7}
\TT
   \textsc{polblogs} & 643 & 2280 & 9 & 9 & 0.04   & 0.04 \\ 
   \textsc{web-google} & 1299 & 2773 & 18 & 18 & 0.01   & 0.06 \\ 
   \textsc{wikipedia} & 1.8M & 4.5M & 31 & 31 & 1.94   & 0.23 \\ 
\bottomrule
\end{tabular}

\end{table}

\begin{figure*}[tb]
 \centering
 \subfigure[global cc vs. max-clique]{\includegraphics[width=0.33\linewidth]{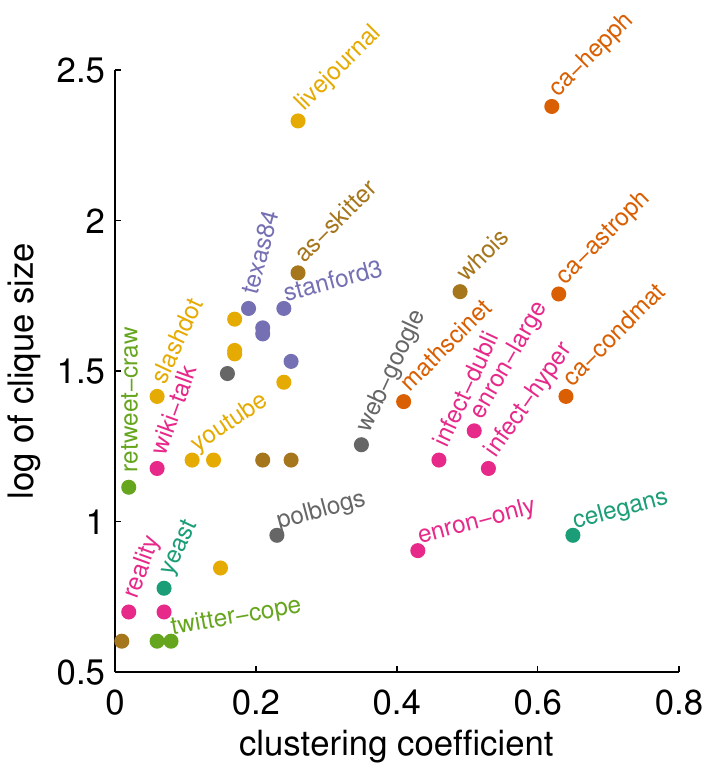}}%
 \subfigure[$d_{\text{avg}}$ vs. max-clique]{\includegraphics[width=0.33\linewidth]{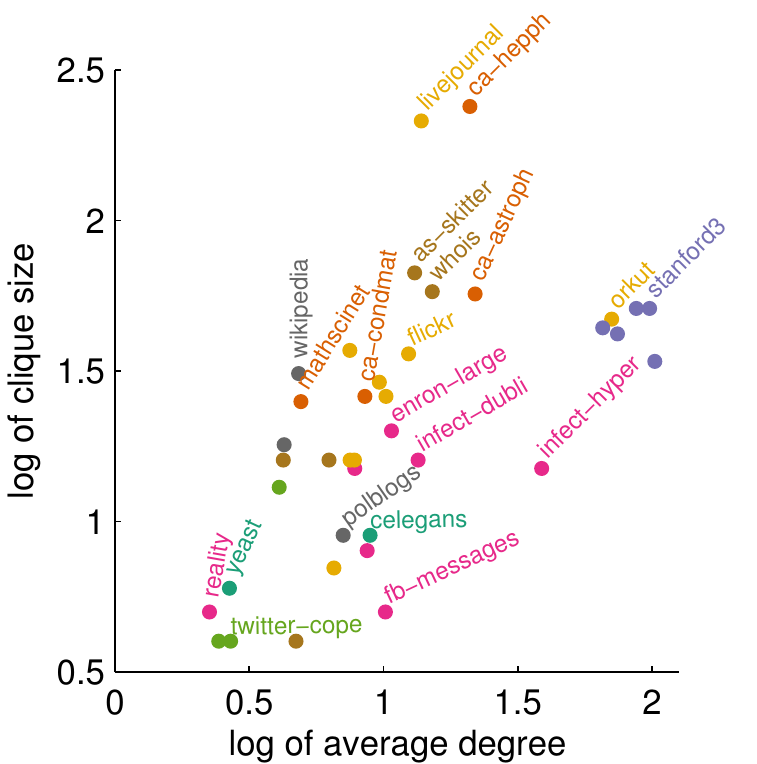}}%
 \subfigure[$t_{\text{avg}}$ vs. max-clique]{\includegraphics[width=0.33\linewidth]{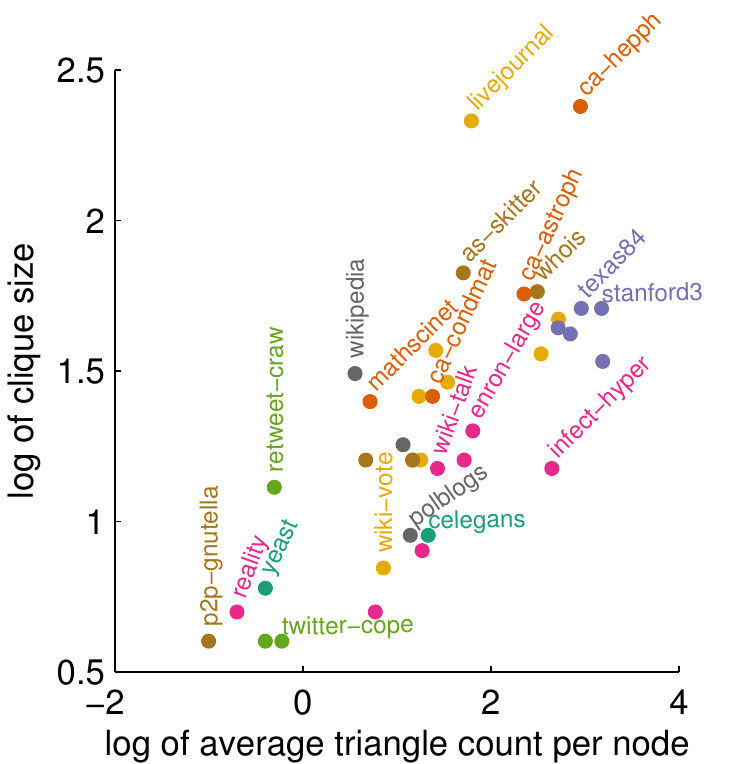}}
 \caption{Each point shows one of the 38 graphs, colored by the graph type. See the discussion in the text.}
 \label{fig:clique-vs-property}
 \end{figure*}

Now, we wish to study the relationships between these clique sizes and various graph properties.  A full table of these results is given in the supplementary material.  Here, we summarize with more informative figures.  We show how the size of the clique varies with the global clustering coefficient (ratio of triangles to wedges), the average degree, and average triangles per node in Figure~\ref{fig:clique-vs-property}.  In the clustering coefficient plot (a), we see little relationship between the clustering coefficient and the maximum clique, whereas we see a much strong relationship between the average number of triangles and clique size (c).  In the average degree plot (b), we also see the set of Facebook networks and the orkut network form a small group.  Among these figures, the strongest relationship with the clique size is the log of the average triangle count.

\begin{figure*}[tb]
\centering
 \includegraphics{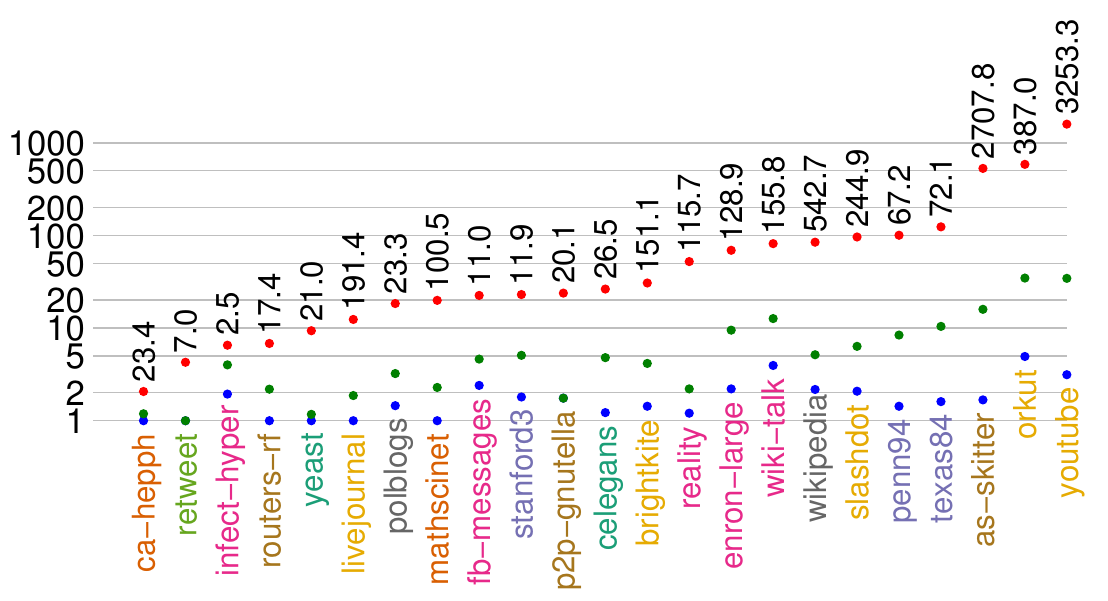}
 \caption{For a subset  of 23 networks, sorted by increasing values of maximum degree over clique size, we show the relationship between the $k$-core upper bound on the clique size (shown as the blue points), the maximum triangle upper-bound on clique size (shown as the green points), and the maximum degree upper-bound on clique size (shown as the red points).  We normalize these by the size of the maximum clique so all values are larger than one.  Thus, we can immediately determine that the $k$-core bound is accurate up to a factor of 5, and usually much more accurate. The upper number is the ratio of maximum degree to average degree -- a measure of the skewness of the degree distribution. }
 \label{fig:bounds}
\end{figure*}

To conclude our analysis of these networks, we study the bounds on the size of the largest clique from the introduction: \[
                \omega(G) -1  \le K(G) \text{ and } \omega(G) - 1 \le \sqrt{2 T(G)}.                                                                                                                          \]
These bounds are shown in Figure~\ref{fig:bounds}. We find that the $k$-core bound is accurate up to a factor of 5, and usually much more accurate.  In the supplementary material, we further include the fraction of the maximum clique contained with the highest $k$-core.  This value is usually high, except for the largest networks.  This finding suggests that as graphs increase in size, the relationship between the $k$-core and maximum clique changes.
 

\section{Temporal Strong Components} \label{sec:tscc}
In this section, we use the maximum clique method as a subroutine to compute temporal strong components~\cite{nicosia:023101}.  Since this area is somewhat new, we review some of the basic definitions in order to motivate the relationship between cliques and temporal strong components.

\begin{Definition}[\small\textsf{temporal network}\normalsize]
Let $V$ be a set of vertices, and $E_T \subseteq V \times V \times \RR^{+}$ be the set of temporal edges between vertices in $V$. Each edge $(u,v,t)$ has a unique time $t \in \RR^{+}$.
\end{Definition}

For such a temporal network, a path represents a sequence of edges that must be traversed in increasing order of edge times.  That is, if each edge represents a contact between two entities, then a path represents a feasible route for a piece of information.
\begin{Definition}[\small\textsf{temporal paths}\normalsize]
A temporal path from $u$ to $w$ in $G=(V,E_T)$ is a sequence of edges $e_1, \ldots, e_k$ such that the $e_1\!=\!(u,v_1,t_1), \ldots, e_k\!=\!(u_k, w, t_k)$ where $v_{j} = u_{j+1}$ and $t_j < t_{j+1}$ for all $j = 1$ to $k$.
We say that $u$ is \textit{temporally connected} to $w$ if there exists a such temporal path.
\end{Definition}
This definition echoes the standard definition of a path, but adds the additional constraint that paths must follow the directionality of time.
Temporal paths are inherently asymmetric because of the directionality of time.

\begin{Definition}[\small\textsf{temporal strong component}\normalsize]
Two vertices $(u,w)$ are strongly connected if there exists a temporal path $\mathcal{P}$ from $u$ to $w$ and from $w$ to $u$.  A temporal strongly connected component (tSCC) is defined as a maximal set of vertices $C \subseteq V$ such that any pair of vertices in $C$ are strongly connected.
\end{Definition}
While the vertices of a strong component in a graph define an equivalency class, and hence, we can partition a network into components, the same fact is not true of temporal strong components.  The vertices in a temporal strong component can overlap with those in another temporal strong component.

Note that a temporal weak component is always equal to the connected component in the static graph~\cite{tang2010characterising}.  We conclude by mentioning that stronger definitions of temporal components exist.  For example, the temporal paths used to define a strong component can be further restricted to only use vertices from the component $C$ itself.

As previously mentioned, checking if a graph has a $k$-node temporal SCC is NP-complete~\cite{bhadra2003complexity,nicosia:023101}.  Nonetheless, we can compute the largest such strong component using a maximum clique algorithm.  Let us briefly explain how.  

The first step is to transform the temporal graph into what is called a strong-reachability graph.  For each pair of vertices in $V$, we place an edge in the strong reachability graph if there is a temporal path between them.  In Algorithm~\ref{alg:tscc}, this corresponds to the \textsc{reach} subroutine.  That method uses the temporal ordering proposed by~\cite{Pan-2011-paths} and builds up temporal paths backwards in time. With this reachability graph, the second step of the computation is to remove any non-reciprocated edges and then find a maximum clique.  That maximum clique is the largest set of nodes where all pairwise temporal paths exist, and hence, is the largest temporal strong component~\cite{nicosia:023101}.

\SetAlgoNoLine
\RestyleAlgo{ruled}
\setlength{\algomargin}{2em}
\SetAlgoCaptionSeparator{}
\SetAlCapHSkip{0pt}
%\IncMargin{-10mm}
\begin{algorithm}[t!]
\caption{Largest Temporal Strong \rlap{Component}}
\label{alg:tscc}
\SetKwInOut{Input}{input}\SetKwInOut{Output}{output}
\KwIn{Dynamic Graph $\mathcal{G} = (V,E_T)$}
\medskip
\textbf{procedure} \textsc{max-tscc}($\mathcal{G} = (V,E_T)$) \\
$E_R = \text{\textsc{reach}}(\mathcal{G})$\\
Remove non-reciprocal edges from $E_R$\\
Compute the \textsc{max-clique} in the graph $(V,E_R)$ \\
Return the subgraph of $G$ induced by $C$
\bigskip

\textbf{procedure} \textsc{reach}($\mathcal{G} = (V,E_T)$) \\
Sort edges to be in reverse time order \\
Set $E_R$ to be the set of all self-loops\\
\For{$(i,j,t) \in E$} 
{
  Add $(i,k)$ to $E_R$ for all $k$ where $(j,k) \in E_R$
}
Return $E_R$
\end{algorithm}

\subsection{Temporal Graph Data.}
We use three types of temporal graph data.  For all networks, we discard self-loops and any edge weights.  In all cases, the nodes represent people.  And in all cases, the largest temporal strong components reflect groups of people that meet, interact, or retweet with each other sufficiently often to transmit any message or meme.

%infectious
\textbf{Contact networks}.
The edges are face-to-face contacts (infect-dublin, infect-hyper\cite{isella2011s}). See ref.~\cite{infect} for more details about these data.

\textbf{Interaction networks}.
The edges are observed interactions such as forum posts (fb-forum~\cite{opsahl2011triadic}) and private messages (fb-messages~\cite{opsahl2009clustering}). 
We also investigate a cellular telephone call network where the edges are calls (reality~\cite{eagle2006reality}). 

\textbf{Retweet networks}.
Here, the edges are retweets.  First, we investigate 26 retweet networks from various social and political hashtags. We collected these from Truthy~\cite{truthy} on September 20th, 2012.  Of these, only a few had non-trivial temporal strong components which we discuss below.  Also, we analyzed a network of political tweets centered around the November 2010 election in the US (retweet~\cite{conover2011political}).
The final dataset is a retweet and mentions network from the UN conference held in Copenhagen. The data was collected over a two week period (twitter-copen~\cite{ahmed2010time}).

\def \lentab{2.5in}
\begin{figure*}[p]
\centering
    \hspace{-10mm}\subfigure[Reality]
    	{\label{fig:reach-infectious}\includegraphics[width=3.7in]{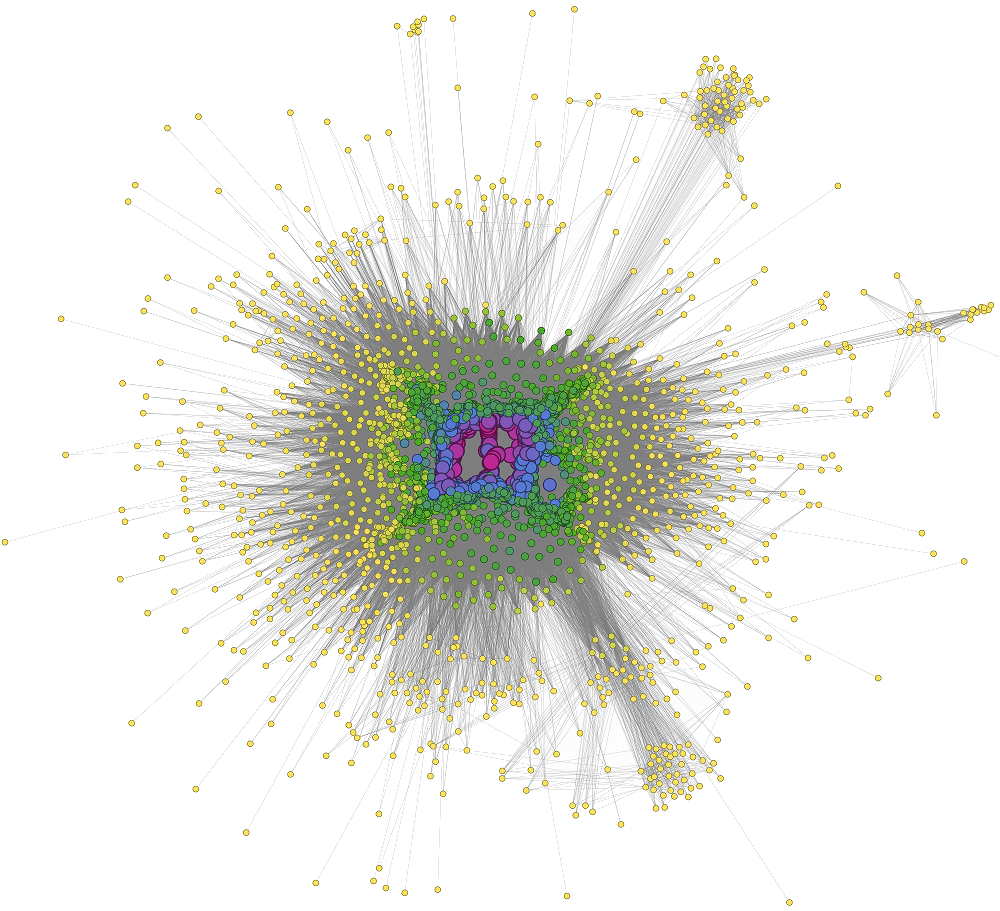}}
    \hspace{-0mm}\subfigure[Retweets (political)]
    	{\label{fig:reach-retweets}\includegraphics[width=3.2in]{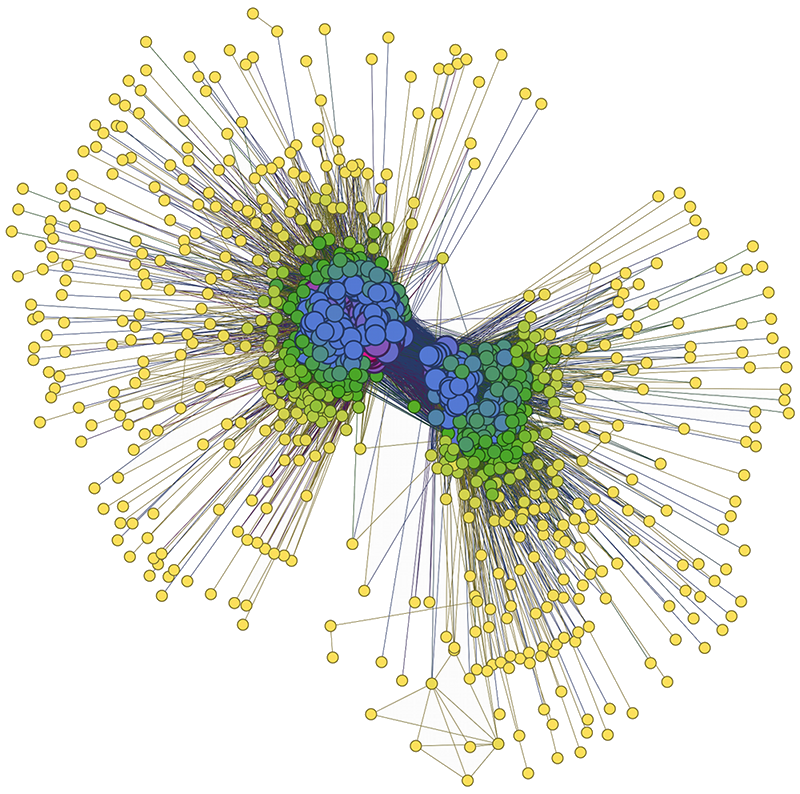}}
  \caption{In order to compute the largest temporal strong component, we first compute the strong reachability network. These networks are rather dense and may reveal clear community structure as in the retweet networks (and, the infect-dublin network, see the supplementary material).}
\label{fig:reach-vis}
    \hspace{0mm}\subfigure[Reality]
    	{\label{fig:tscc-infectious}\includegraphics[width=3.2in]{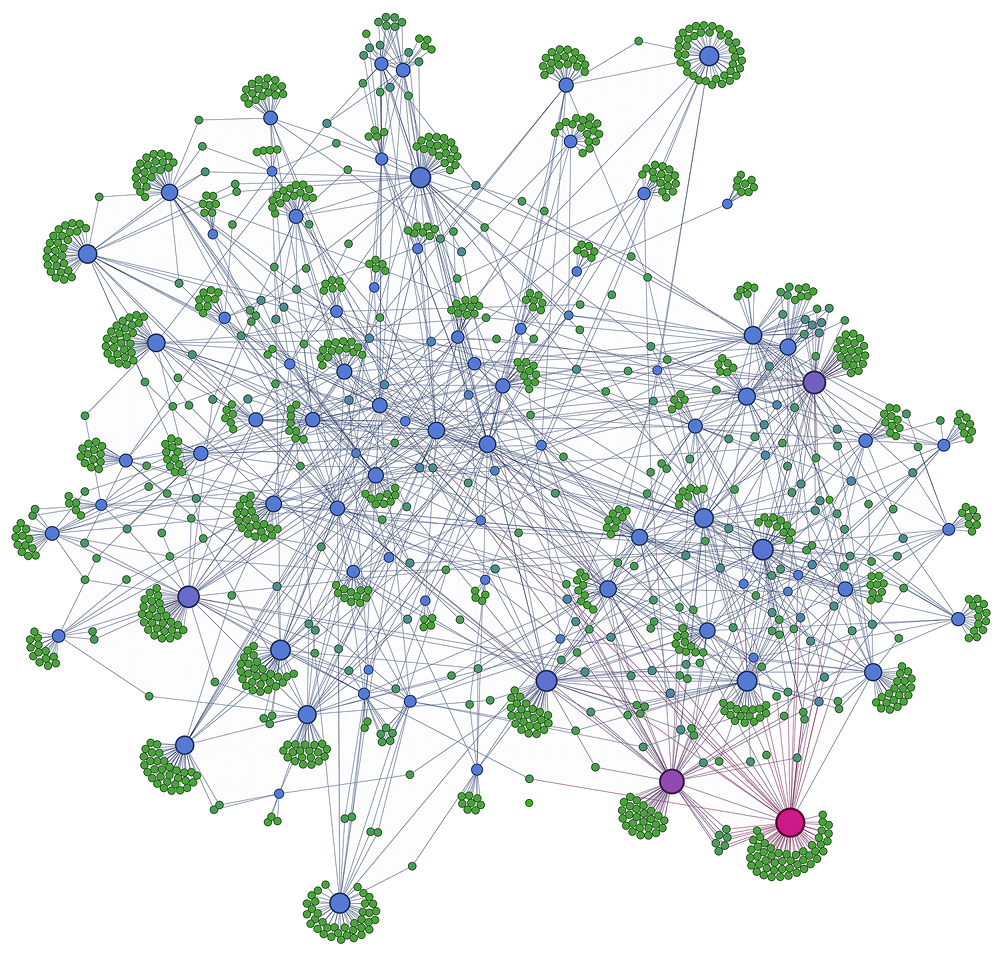}}
    \hspace{-0mm}\subfigure[Retweets (political)]
    	{\label{fig:tscc-retweets}\includegraphics[width=3.4in]{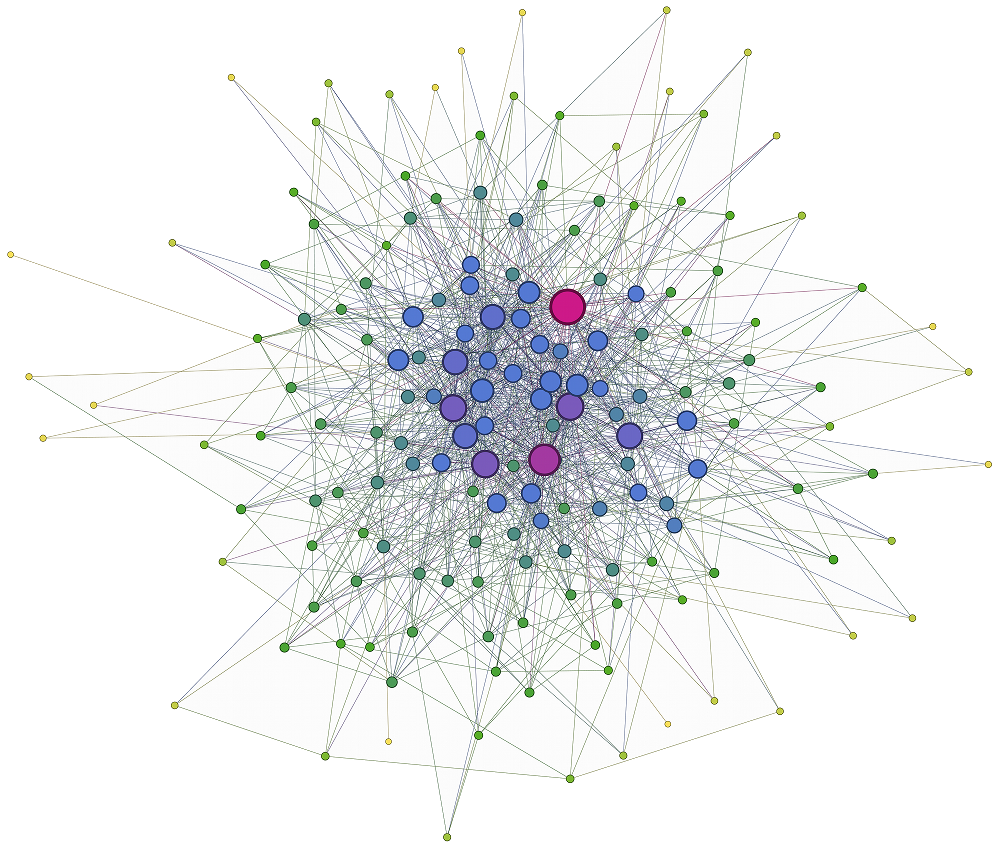}}
    	\vspace{-3mm}
  \caption{
\emph{(a)} In the largest$^\ast$ (see discussion) temporal strong component of the reality network, we see a small group of core users maintaining connectivity among various groups, \emph{(b)} whereas the largest$^\ast$  component in the retweet network consists of 166 twitter users classified as politically ``right'' according to the original data with only a single exception.  These users are a subset of one of the communities.}
  \label{fig:tscc-vis}
\end{figure*}

\subsection{Results and analysis}

We now discuss the results of running our temporal strong component algorithms.  Figures~\ref{fig:reach-vis}~and~\ref{fig:tscc-vis} show the reachability and largest temporal strong component from two networks, respectively.  For the both networks, the exact clique finder did not complete within the specified period.  We mark that fact with an ``$\ast$'' to denote that it may not be the largest. The heuristic procedure found a clique with 1217 vertices for the reality network and 166 vertices for the retweets network.  Given that the largest cores in these networks are 1236 and 175, respectively, in the worst case, we are only off by a few vertices. Our finding about the political party for the retweets also holds for the $166$-core of the network, so we can be confident there is not a larger temporal strong component on the ``political left.''

We summarize the remaining experiments on the temporal strong components in Tables~\ref{table:retweet-dynamic}~and~\ref{table:tscc}.  The first table presents basic network statistics of the original temporal graphs $(G)$ as well as their strong reachability graphs $(\mR_s)$.  For both cases, we list the largest clique we computed with either the exact or heuristic clique finder.  Like above, those cliques marked with a ``$\ast$'' represent heuristic computations because we only let the exact clique algorithm run for one hour.  However, even for these heuristics, we find that they are strikingly close to the upper-bound on the clique size given by the $k$-core.  In fact, in all of the other examples, the largest clique in these networks \emph{was} the largest $k$-core.  Thus, just as in the political retweet network above, we are confident that our results are close to the true largest component.  

The second table (Table~\ref{table:tscc}) shows the time required for the maximum clique routine to return the largest temporal strong component -- the largest clique in the reachability graph.  As mentioned previously, we terminated the exact clique algorithm after an hour.  In this table, we also include statistics on the largest strong component for the hashtag networks from Truthy.  These show that there are heavily interacting groups of tweeters in these datasets that we can identify quickly (less than a second) using our methods.  See the supplementary material for the network statistics on these hashtag networks.  

Overall, we find a few interesting properties of these temporal strong components. In the contact networks (infect-hyper and infect-dublin), both of the largest strong components had about 100 vertices, despite the drastically different sizes of the initial dataset. We suspect this is a property of the data collection methodology because the infect-dublin data were collected over months, instead of days for the infect-hyper.  In the interaction networks, the temporal components contain a significant fraction of the total vertices, roughly 20-30\%.  Finally, in the retweet networks, these temporal strong components are around 20\% of the total vertices in the reachability graph.  Given the strong communication pattern between these groups, they are good candidates for the center of a community in these networks.

\begin{table*}[t!]
\caption{Network statistics of the dynamic networks and their strong reachability graphs. The cases where the max k-core gives the exact size of the max clique are highlighted in gray.  Here, $d_{\max}$ and $d_{\text{avg}}$ are the largest and average degrees, respectively.  Also, $\bar{\kappa}$ is the mean clustering coefficient; $T$ and $T_{\text{avg}}$ are the maximum number and average number of triangles incident on a vertex; $K$ is the largest degree for a $k$-core to exist, and $\omega$ is the clique number (or a lower-bound from the heuristic if the value has an $\ast$).}
\label{tab:misc-dynamic}
\label{table:retweet-dynamic}

\centering
\footnotesize

\begin{tabularx}{\linewidth}{@{} ll XXXXXX XXXX@{}}
\toprule
\textbf{Graph} & & $|V|$ & $|E|$ & 
$d_{\max}$ & $d_{\text{avg}}$ & $\mathbf{\bar{\kappa}}$ & 
$T$ & $T_{\text{avg}}$ & $\sqrt{2T}$ & $K$ & $\omega$ \\
\midrule
% \multirow{2}{*}{\rotatebox{0}{\textsc{enron-only}}}
%  	 & $\rm G$ & 151 & 50572 & 74 & 14.2 & 0.45 & 246 & 47 & 22 & 11 & \textbf{8}\\ 
% 	 & $\mR_s$ & 146 & 9828 & 145 & 134.6 & 0.96 & 9683 & 8738 & 139 & \cellcolor{lightgray} 119 & \cellcolor{lightgray} \textbf{120}\\ 
% \midrule 
\multirow{2}{*}{\rotatebox{0}{\textsc{infect-hyper}}}
 	 & $\rm G$ & 113 & 20818 & 98 & 38.9 & 0.53 & 1731 & 448 & 59 & 28 & \textbf{15}\\ 
	 & $\mR_s$ & 113 & 6222 & 112 & 110.1 & 0.99 & 6110 & 5934 & 111 & \cellcolor{lightgray} 105 & \cellcolor{lightgray} \textbf{106}\\ 
\midrule 
\multirow{2}{*}{\rotatebox{0}{\textsc{infect-dublin}}}
 	 & $\rm G$ & 10972 & 415912 & 64 & 8.1 & 0.45 & 315 & 22 & 25 & 18 & \textbf{16}\\ 
	 & $\mR_s$ & 10972 & 175573 & 219 & 32.0 & 0.80 & 9129 & 633 & 135 & \cellcolor{lightgray} 83 & \cellcolor{lightgray} \textbf{84}\\ 
\midrule
\multirow{2}{*}{\rotatebox{0}{\textsc{reality}}}
 	 & $\rm G$ & 6809 & 52050 & 261 & 2.3 & 0.02 & 52 & 0 & 10 & 5 & \textbf{5}\\ 
	 & $\mR_s$ & 6809 & 4714485 & 6069 & 1384.8 & 0.87 & 4695661 & 971909 & 3065 & 1235 & \textbf{1217}\rlap{$^\ast$}\\ 
\midrule 
	 \multirow{2}{*}{\rotatebox{0}{\textsc{retweet}}} 
& $\rm G$ & 18470 & 61157 & 785 & 2.6 & 0.01 & 238 & 1 & 22 & 5 & \textbf{3}\\ 
& $\mR_s$ & 1206 & 65990 & 551 & 109.4 & 0.70 & 52407 & 9073 & 324 & 174 & \textbf{166}\\ 
\midrule 
\multirow{2}{*}{\rotatebox{0}{\textsc{twitter-copen}}}
 	  	 & $\rm G$ & 8581 & 45933 & 215 & 3.4 & 0.07 & 1210 & 3 & 49 & 5 & \textbf{4}\\ 
	 & $\mR_s$ & 2623 & 473614 & 1516 & 361.1 & 0.66 & 469039 & 111007 & 969 & 582 & \textbf{573}\rlap{$^\ast$}\\ 
% \multirow{2}{*}{\rotatebox{0}{\textsc{retweet-craw}}}
%  	 & $\rm G$ & 1132039 & 4095614 & 5070 & 4.0 & 0.02 & 1555 & 0 & 56 & 18 & \textbf{13}\\ 
% 	 & $\mR_s$ & 17151 & 24015 & 195 & 2.8 & 0.27 & 1056 & 4 & 46 & \cellcolor{lightgray} 19 & \cellcolor{lightgray} \textbf{20}\\ 
\midrule 
%\rowcolor{lightblue}
% \multirow{2}{*}{\rotatebox{0}{\textsc{retweet-crawl}}}
%      & $\rm G$ & 1132039 & 4095614 & 5070 & 4.0 & 0.02 & 1555 & 0 & 39 & 18 & \textbf{13}\\ 
%  & $\mR_s$ & 17151 & 24015 & 195 & 2.8 & 0.27 & 1056 & 4 & 32 & \cellcolor{lightgray}	19 & \cellcolor{lightgray} \textbf{20}\\ 
% \midrule 
\multirow{2}{*}{\rotatebox{0}{\textsc{fb-forum}}}
 	 & $\rm G$ & 899 & 33720 & 128 & 15.7 & 0.06 & 407 & 16 & 29 & 14 & \textbf{3}\\ 
	 & $\mR_s$ & 488 & 71011 & 435 & 291.0 & 0.86 & 70402 & 44845 & 375 & 272 & \textbf{262}\rlap{$^\ast$}\\ 
\midrule 
\multirow{2}{*}{\rotatebox{0}{\textsc{fb-messages}}}
 	 & $\rm G$ & 1899 & 61734 & 255 & 14.6 & 0.11 & 1095 & 23 & 47 & 20 & \textbf{5}\\ 
	 & $\mR_s$ & 1303 & 531893 & 1253 & 816.4 & 0.85 & 530538 & 317651 & 1030 & 706 & \textbf{677}\rlap{$^\ast$}\\ 
\midrule 
\end{tabularx}

\end{table*}

\begin{table}[t!]
\caption{ For each of the temporal networks, we list the largest clique found by the exact method (if it finished in an hour) or via the heuristic, as well as the time taken by each algorithm.  The exact method completes quickly for the contact networks, and hashtag networks, but has trouble with the interaction networks and retweet networks.  In these cases, the heuristic gives a good lower-bound, that isn't too far away from the $k$-core upper-bound.}
\label{table:tscc}
\centering\small
\begin{tabularx}{\columnwidth}{l@{\,\,\,}XXXX}
\toprule
& & & \multicolumn{2}{l}{Time (sec.)} \\
Graph & $\omega$ & $\tilde{\omega}$ & Exact & Heur. \\
\midrule
   \textsc{infect-dublin} & 84 & 84 & 4.73  & 0.1\\ 
   \textsc{infect-hyper} & 106 & 105 & 0.03 & 0.05\\ 
   %\textsc{enron-only} & 120 & 116 & 807.07 & 0.08\\ 
\midrule
   \textsc{fb-forum} & $\le 273$ & 262 & $>3600$  & 1\\ 
   \textsc{fb-messages} & $\le 706$  & 677 & $>3600$  & 46.31\\    
   \textsc{reality} & $\le 1236$ & 1217 & $>3600$  & 1039.55\\ 
\midrule
   %\textsc{retweet-craw} & 20 & 17 & 0.07 s  & 0.11 s\\ 
   \textsc{retweet} & $\le 175$ & 166 & $>3600$  & 0.82\\ 
   \textsc{twitter-copen} & $\le 582$ & 573 & $>3600$  & 26.71\\ 
\midrule   
   \textsc{\#assad} & 8 & 7 & $<0.1$  & $<0.1$\\ 
   \textsc{\#bahrain} & 8 & 6 & $<0.1$  & $<0.1$\\ 
   \textsc{\#saudi} & 8 & 8 & $<0.1$  & $<0.1$\\ 
   \textsc{\#barackobama} & 10 & 10 & $<0.1$  & $<0.1$\\ 
   \textsc{\#alwefaq} & 16 & 16 & $<0.1$  & $<0.1$\\ 
   \textsc{\#justinbieber} & 17 & 16 & $<0.1$  & $<0.1$\\ 
   \textsc{\#occupywallst} & 18 & 18 & $<0.1$  & $<0.1$\\ 
   \textsc{\#gmanews} & 22 & 21 & $<0.1$  & $<0.1$\\ 
   \textsc{\#onedirection} & 27 & 27 & $<0.1$  & $<0.1$\\ 
   \textsc{\#lolgop} & 42 & 40 & 0.1  & $<0.1$\\ 
\bottomrule
\end{tabularx}

\end{table}

\section{Maximum Clique Algorithms}
\label{sec:clique-algs}
We now provide some algorithmic details of  the maximum clique finders 
we have used in the previous two sections.
We first outline the algorithms as introduced in 
\cite{patt2012cliques} and then discuss the extensions 
we made in this work to improve performance and utility.

\subsection{The Exact Algorithm}
Given an undirected graph $G=(V,E)$,  let $C_v$ denote a clique of the largest size 
containing the vertex $v$. 
A maximum clique in  $G$ can be found by computing $C_v$ for
every vertex $v$ in $V$ and then picking the largest  among these.
The exact algorithm we use follows this basic strategy, 
but drastically speeds up the process by avoiding the computation of every
$C_u$, $u \in V$,  that would eventually be smaller than the global maximum.
For such {\em pruning} purposes, the algorithm 
maintains  the size of  the largest clique found so far  ($MaxSoFar$). 

For a given vertex $v$,  $C_v$ is evaluated by performing
a {\em depth-first search} starting from $v$. 
At each {\em depth} of the search, the algorithm compares 
$MaxSoFar$ against the number of remaining
vertices that could potentially constitute a clique containing the vertex $v$.
If the latter number is found to be smaller, 
the algorithm backtracks (the search is pruned).
The algorithm employs several other pruning steps in addition to the backtracks.
For example, a depth-first search from a vertex $v$ is invoked only if
the degree of $v$ is greater than or equal to $MaxSoFar$, 
since otherwise  the largest clique containing $v$ would certainly be smaller than 
$MaxSoFar$. 

\subsection{The Heuristic}

In its search for a maximum clique in the entire graph,  
the exact algorithm described explores for every vertex $v$ 
feasible cliques  along a subset of the edges incident on $v$.
The heuristic instead picks just {\em one} of these edges for further
exploration. The picked edge is one incident on a vertex
with the {\em largest degree} among the neighbors of $v$.
The rationale for this choice is that the neighbor of $v$ with the largest
degree is more likely than other neighbors of $v$ to be part of the 
largest clique on $v$ ($C_v$). 
The exact algorithm in the worst case has exponential runtime,
whereas the runtime of the heuristic is bounded by $O(|V|\cdot \Delta^2)$,
where $\Delta$ is the maximum degree in the input graph.

\subsection{New Extensions}

In this work, we extended both the exact algorithm and the heuristic in two major ways.

First, we {\em parametrized} the exact algorithm such that it takes 
both a lower bound $lb$ and an upper bound $ub$ on the size of the clique
to be computed. The algorithm starts by setting $MaxSoFar$ to $lb$---thus a good lower bound
helps speed up the algorithm by allowing early pruning. And the algorithm stops
the search as soon as a clique of size equal to $ub$ is found---in the absence
of a bound, the algorithm may continue to iterate over the remaining vertices even 
if the global maximum clique had already been found. 

Second, we {\em parallelized} the algorithms.
The computations of  per-vertex maximum cliques ($C_v$'s) over the vertices are 
independent of each other---and hence can be done in parallel---except for one issue: 
the need for information on the size of the global maximum clique found so far. 
However, since this information is needed only for pruning purposes, one need not 
insist on disseminating it to all processing units as soon as it is updated.
Instead, one could work with a locally computed maximum clique for pruning purposes
and occasionally exchange the global value at regular synchronization points.
Following this strategy, we parallelized, using OpenMP threading, both the 
exact algorithm and the heuristic. 

\begin{figure}[t]
\includegraphics[width=3in]{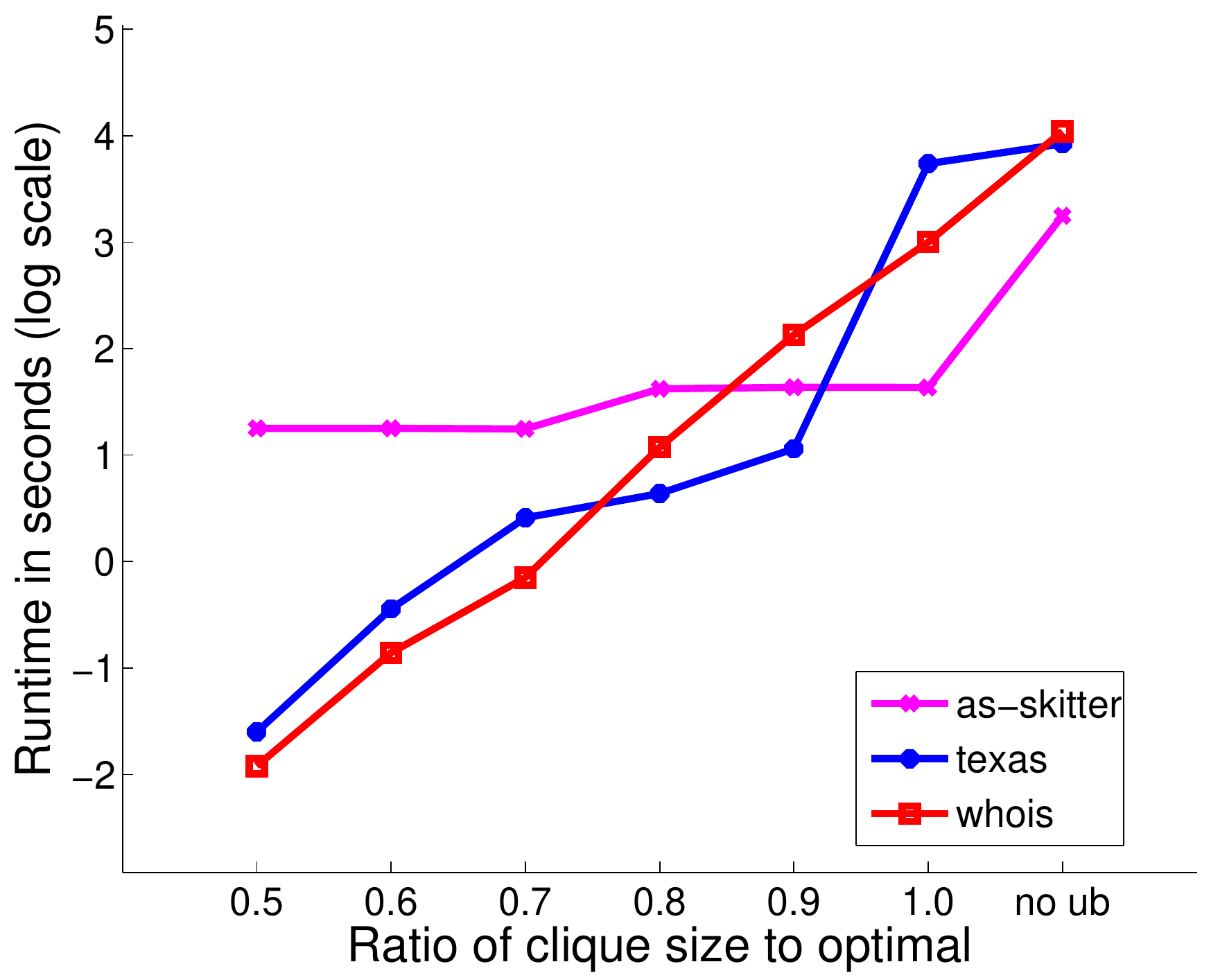}
\caption{Runtime behavior as upper bound on the clique algorithm is varied.}
\label{fig:cliqueTimeAnalysis}
\end{figure}

The first improvement, parameterization, makes possible added flexibility.
By intentionally setting $ub$ to be smaller than a true upper bound on a maximum
clique, a user may force the algorithm to terminate when a clique of the desired size 
($ub$) is found. Figure~\ref{fig:cliqueTimeAnalysis} illustrates the benefit of this.
It shows how the runtime of the algorithm varies as the $ub$ is varied for three 
large networks from our dataset ({\tt as-skitter}, {\tt texas}, and {\tt whois}).
Note that, in general,  the algorithm detects cliques of small sizes extremely fast:
in fact cliques of up to $90\%$ of the maximum clique size
can be computed at under $5\%$ of the time it takes to compute the maximum clique. 
The plots also provide some insight into the different scenarios the three networks 
represent: in the {\tt as-skitter} network, computing a clique of one size 
takes nearly the same time as computing a clique of just about any other size---meaning that
cliques are easy to find---whereas in the other two networks it is increasingly difficult
to detect larger cliques.

%The default value for $lb$ is zero, and that for $ub$ is  the largest $k$ 
%for which there is a $k$-core in the graph---this value is
%computed in the clique algorithm itself. 

\section{Discussion}

In the title, we hypothetically asked ``What if \textsc{clique} were fast?''  We found that a state-of-the-art clique finder can identify maximum cliques in sparse networks with hundreds of millions of edges in a few seconds, but in the worst case, a few hours. These results show that cliques in social and information networks are not always difficult to find and hint at a deeper structural property of these networks. The same maximum clique finder was less successful at identifying maximum cliques in temporal reachability networks (which give rise to temporal strong components).  These networks have significant structure implied by the temporal edges that are hidden from the clique finder.  This structure and the density of these networks may make these problems especially challenging.  In the future, we plan to investigate additions to the clique finder to exploit such structure.

\newpage
\bibliographystyle{abbrv}
\bibliography{rossi-tscc,assefaw-cliques,gleich}

%\subsection{Data}
%We briefly discuss the types of \textit{static graphs} used in our empirical investigation.
%For these graphs, we only look at the largest connected component.
%There are X types of graphs.
%
%\textbf{Biological networks.}
%
%\textbf{Collaboration networks.}
%
%\textbf{Social networks.}
%
%\textbf{Technological networks.}
%
%\textbf{Web graphs.}
%
%\textbf{Contact networks.}

\begin{table*}[b!]
\section*{Appendix}
\caption{Network statistics of the static graphs from \textbf{section 2}, including the size of the maximum clique and corresponding upperbounds. 
% The cases where the max k-core gives the exact size of the max clique are highlighted in gray.  
Here, $d_{\max}$ and $d_{\text{avg}}$ are the largest and average degrees, respectively.  Also, $\bar{\kappa}$ is the mean clustering coefficient; $T$ and $T_{\text{avg}}$ are the maximum number and average number of triangles incident on a vertex; $K$ is the largest degree for a $k$-core to exist, $\omega$ is the clique number, and $\Gamma_K = \frac{|C \cap K|}{|C|}$ is the recall measurement by using the $k$-core to find the maximum clique.}
%\caption{Properties of the static graphs from \textbf{section 2}, including the size of the max clique and corresponding upperbounds. We denote $\Gamma_K$ as the max clique recall from the max kcore: $\Gamma_K = \frac{|C \cap K|}{|C|}$.}
\vspace{1mm}
\label{table:mc-data}
\centering\small\scriptsize
\begin{tabularx}{\linewidth}{r@{\,\,\,}XXXXXXXXXXX}
\toprule
\textbf{Graph} & $|V|$ & $|E|$ & 
$d_{\max}$ & $d_{\text{avg}}$ & $\mathbf{\bar{\kappa}}$ & 
$T$ & $T_{\text{avg}}$ & $\sqrt{2T}$ & $K$ & $\omega$ &  $\Gamma_K$\\
\midrule
   \textsc{yeast} & 1458 & 1948 & 56 & 2.7 & 0.07 & 18 & 0.4 & 6 & 5 & \textbf{6} & 1.00 \\ 
   \textsc{celegans} & 453 & 2025 & 237 & 8.9 & 0.65 & 870 & 21.7 & 42 & 10 & \textbf{9} & 0.89 \\ 
\midrule 
   \textsc{mathscinet} & 332689 & 820644 & 496 & 4.9 & 0.41 & 1564 & 5.2 & 56 & 24 & \textbf{25} & 1.00 \\ 
   \textsc{ca-condmat} & 21363 & 91286 & 279 & 8.5 & 0.64 & 1615 & 24.0 & 57 & 25 & \textbf{26} & 1.00 \\ 
   \textsc{ca-astroph} & 17903 & 196972 & 504 & 22.0 & 0.63 & 11269 & 226.2 & 150 & 56 & \textbf{57} & 1.00 \\ 
   \textsc{ca-hepph} & 11204 & 117619 & 491 & 21.0 & 0.62 & 39633 & 899.1 & 282 & 238 & \textbf{239} & 1.00 \\ 
\midrule 
   \textsc{fb-messages} & 1266 & 6451 & 112 & 10.2 & 0.07 & 242 & 5.9 & 22 & 11 & \textbf{5} & 1.00 \\ 
   \textsc{reality} & 6809 & 7680 & 261 & 2.3 & 0.02 & 52 & 0.2 & 10 & 5 & \textbf{5} & 0.80 \\ 
   \textsc{enron-only} & 143 & 623 & 42 & 8.7 & 0.43 & 125 & 18.7 & 16 & 9 & \textbf{8} & 1.00 \\ 
   \textsc{infect-hyper} & 113 & 2196 & 98 & 38.9 & 0.53 & 1731 & 447.8 & 59 & 28 & \textbf{15} & 0.93 \\ 
   \textsc{wiki-talk} & 92117 & 360767 & 1220 & 7.8 & 0.06 & 17699 & 27.2 & 188 & 58 & \textbf{15} & 1.00 \\ 
   \textsc{infect-dubli} & 410 & 2765 & 50 & 13.5 & 0.46 & 280 & 52.1 & 24 & 17 & \textbf{16} & 1.00 \\ 
   \textsc{enron-large} & 33696 & 180811 & 1383 & 10.7 & 0.51 & 17744 & 64.6 & 188 & 43 & \textbf{20} & 0.80 \\ 
%\midrule 
%   \textsc{amazon} & 91813 & 125704 & 5 & 2.7 & 0.27 & 9 & 1.1 & 4 & 4 & \textbf{5} & 1.00 \\ 
%\midrule 
%   \textsc{roadnet-ca} & 1957027 & 2760388 & 12 & 2.8 & 0.05 & 7 & 0.2 & 4 & 3 & \textbf{4} & 1.00 \\ 
%   \textsc{roadnet-pa} & 1087562 & 1541514 & 9 & 2.8 & 0.05 & 8 & 0.2 & 4 & 3 & \textbf{4} & 1.00 \\ 
\midrule 
   \textsc{retweet} & 96 & 117 & 17 & 2.4 & 0.06 & 6 & 0.4 & 3 & 3 & \textbf{4} & 1.00 \\ 
   \textsc{twitter-cope} & 761 & 1029 & 37 & 2.7 & 0.08 & 27 & 0.6 & 7 & 4 & \textbf{4} & 1.00 \\ 
   \textsc{retweet-craw} & 1112702 & 2278852 & 5070 & 4.1 & 0.02 & 1555 & 0.5 & 56 & 18 & \textbf{13} & 0.00 \\ 
\midrule 
   \textsc{wiki-vote} & 889 & 2914 & 102 & 6.6 & 0.15 & 251 & 7.2 & 22 & 9 & \textbf{7} & 0.57 \\ 
   \textsc{epinions} & 26588 & 100120 & 443 & 7.5 & 0.14 & 5151 & 18.0 & 101 & 32 & \textbf{16} & 0.00 \\ 
   \textsc{youtube} & 495957 & 1936748 & 25409 & 7.8 & 0.11 & 151081 & 14.8 & 550 & 49 & \textbf{16} & 0.94 \\ 
   \textsc{slashdot} & 70068 & 358647 & 2507 & 10.2 & 0.06 & 13382 & 17.2 & 164 & 53 & \textbf{26} & 1.00 \\ 
   \textsc{gowalla} & 196591 & 950327 & 14730 & 9.7 & 0.24 & 93817 & 34.7 & 433 & 51 & \textbf{29} & 0.00 \\ 
   \textsc{flickr} & 513969 & 3190452 & 4369 & 12.4 & 0.17 & 524525 & 343.0 & 1024 & 309 & \textbf{35}$\ast$ & - \\ 
   \textsc{brightkite} & 56739 & 212945 & 1134 & 7.5 & 0.17 & 11517 & 26.1 & 152 & 52 & \textbf{37} & 1.00 \\ 
   \textsc{orkut} & 2997166 & 106349209 & 27466 & 71.0 & 0.17 & 1313133 & 525.1 & 1621 & 230 & \textbf{47} & 0.00 \\ 
   \textsc{livejournal} & 4033137 & 27933062 & 2651 & 13.9 & 0.26 & 79740 & 62.1 & 399 & 213 & \textbf{214} & 1.00 \\ 
\midrule 
   \textsc{duke14} & 9885 & 506437 & 1887 & 102.5 & 0.25 & 41982 & 1560.7 & 290 & 85 & \textbf{34} & 1.00 \\ 
   \textsc{berkeley13} & 22900 & 852419 & 3434 & 74.4 & 0.21 & 69511 & 703.4 & 373 & 64 & \textbf{42} & 1.00 \\ 
   \textsc{penn94} & 41536 & 1362220 & 4410 & 65.6 & 0.21 & 68097 & 520.6 & 369 & 62 & \textbf{44} & 1.00 \\ 
   \textsc{stanford3} & 11586 & 568309 & 1172 & 98.1 & 0.24 & 33177 & 1511.1 & 258 & 91 & \textbf{51} & 1.00 \\ 
   \textsc{texas84} & 36364 & 1590651 & 6312 & 87.5 & 0.19 & 141050 & 922.2 & 531 & 81 & \textbf{51} & 1.00 \\ 
\midrule 
   \textsc{p2p-gnutella} & 62561 & 147878 & 95 & 4.7 & 0.01 & 17 & 0.1 & 6 & 6 & \textbf{4} & 0.00 \\ 
   \textsc{internet-as} & 40164 & 85123 & 3370 & 4.2 & 0.21 & 8513 & 4.7 & 130 & 23 & \textbf{16} & 1.00 \\ 
   \textsc{routers-rf} & 2113 & 6632 & 109 & 6.3 & 0.25 & 588 & 14.8 & 34 & 15 & \textbf{16} & 1.00 \\ 
   \textsc{whois} & 7476 & 56943 & 1079 & 15.2 & 0.49 & 22271 & 314.0 & 211 & 88 & \textbf{58} & 1.00 \\ 
   \textsc{as-skitter} & 1694616 & 11094209 & 35455 & 13.1 & 0.26 & 564609 & 50.9 & 1063 & 111 & \textbf{67} & 0.00 \\ 
\midrule 
   \textsc{polblogs} & 643 & 2280 & 165 & 7.1 & 0.23 & 392 & 14.0 & 28 & 12 & \textbf{9} & 1.00 \\ 
   \textsc{web-google} & 1299 & 2773 & 59 & 4.3 & 0.35 & 189 & 11.7 & 19 & 17 & \textbf{18} & 1.00 \\ 
   \textsc{wikipedia} & 1864433 & 4507315 & 2624 & 4.8 & 0.16 & 12404 & 3.6 & 158 & 66 & \textbf{31} & 0.00 \\ 
\midrule 
\end{tabularx} 
\end{table*}

\begin{table*}
\caption{Network statistics of the \textbf{dynamic retweet graphs} and their strong reachability graphs from \textbf{section 3}.}
\vspace{1mm}
\label{table:reach-data}
\centering\small
\begin{tabularx}{\linewidth}{ ll XXXXX XXc c}
\toprule
%\textbf{graphs} & & \textbf{verts} & \textbf{edges} & 
%$\mathbf{max \; d}$ & $\mathbf{avg \; d}$ & $\mathbf{\bar{\kappa}}$ & 
%$\mathbf{max \; t}$ & $\tau$ & 
%$\mathbf{max}$ \textbf{k-core} & $\mathbf{max} \; \mathlarger{\mathlarger{\mathit{\mathbf{clique}}}}$\\
\textbf{Graph} & & $|V|$ & $|E|$ & 
$d_{\max}$ & $d_{\text{avg}}$ & $\mathbf{\bar{\kappa}}$ & 
$T$ & $\sqrt{2T}$ & $K$ & $\omega$\\
\midrule
\multirow{2}{*}{\rotatebox{0}{\textsc{\scriptsize \#occupywal}}}
 	 & $\rm G$ & 3609 & 3936 & 2811 & 1.1 & 0.02 & 8 & 3 & 1 & \textbf{2}\\ 
	 & $\mR_s$ & 127 & 931 & 112 & 14.7 & 0.72 & 809 & 28 & 17 & \textbf{18}\\ 
\midrule 
\multirow{2}{*}{\rotatebox{0}{\textsc{\scriptsize \#alwefaq}}}
 	 & $\rm G$ & 4171 & 7123 & 865 & 1.7 & 0.07 & 77 & 9 & 4 & \textbf{3}\\ 
	 & $\mR_s$ & 72 & 355 & 39 & 9.9 & 0.50 & 280 & 17 & 15 & \textbf{16}\\ 
\midrule 
\multirow{2}{*}{\rotatebox{0}{\textsc{\scriptsize \#bahrain}}}
 	 & $\rm G$ & 4676 & 8007 & 256 & 1.7 & 0.01 & 15 & 4 & 2 & \textbf{3}\\ 
	 & $\mR_s$ & 72 & 129 & 20 & 3.6 & 0.28 & 66 & 8 & 7 & \textbf{8}\\ 
\midrule 
\multirow{2}{*}{\rotatebox{0}{\textsc{\scriptsize \#mittromne}}}
 	 & $\rm G$ & 7974 & 8597 & 1282 & 1.1 & 0.00 & 16 & 4 & 1 & \textbf{2}\\ 
	 & $\mR_s$ & 102 & 108 & 25 & 2.1 & 0.20 & 38 & 6 & 4 & \textbf{5}\\ 
\midrule 
\multirow{2}{*}{\rotatebox{0}{\textsc{\scriptsize \#gmanews}}}
 	 & $\rm G$ & 8373 & 8832 & 6146 & 1.1 & 0.00 & 98 & 10 & 1 & \textbf{2}\\ 
	 & $\mR_s$ & 135 & 1078 & 114 & 16.0 & 0.72 & 950 & 31 & 21 & \textbf{22}\\ 
\midrule 
\multirow{2}{*}{\rotatebox{0}{\textsc{\scriptsize \#justinbie}}}
 	 & $\rm G$ & 9405 & 9615 & 3176 & 1.0 & 0.00 & 2 & 1 & 1 & \textbf{2}\\ 
	 & $\mR_s$ & 62 & 442 & 42 & 14.3 & 0.69 & 375 & 19 & 16 & \textbf{17}\\ 
\midrule 
\multirow{2}{*}{\rotatebox{0}{\textsc{\scriptsize \#barackoba}}}
 	 & $\rm G$ & 9631 & 9826 & 6107 & 1.0 & 0.00 & 5 & 2 & 1 & \textbf{2}\\ 
	 & $\mR_s$ & 80 & 226 & 43 & 5.7 & 0.48 & 153 & 12 & 9 & \textbf{10}\\ 
\midrule 
\multirow{2}{*}{\rotatebox{0}{\textsc{\scriptsize \#lolgop}}}
 	 & $\rm G$ & 9765 & 10324 & 8280 & 1.1 & 0.01 & 8 & 3 & 1 & \textbf{2}\\ 
	 & $\mR_s$ & 273 & 4510 & 249 & 33.0 & 0.73 & 4247 & 65 & 41 & \textbf{42}\\ 
\midrule 
\end{tabularx} 
\end{table*}

\begin{figure*}[p]
\centering
   \hspace{0mm}\subfigure[Reachability Graph]
    {\label{fig:reach-infectious}\includegraphics[width=3in]{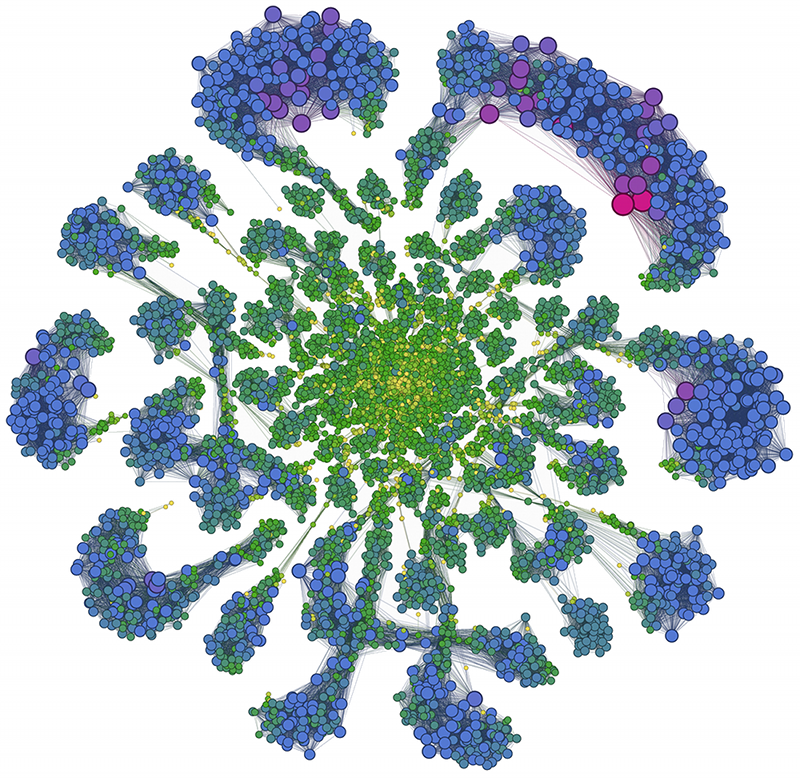}}    
       \hspace{0mm}\subfigure[Largest Strong Temporal Component]
   {\label{fig:tscc-infectious}\includegraphics[width=2.8in]{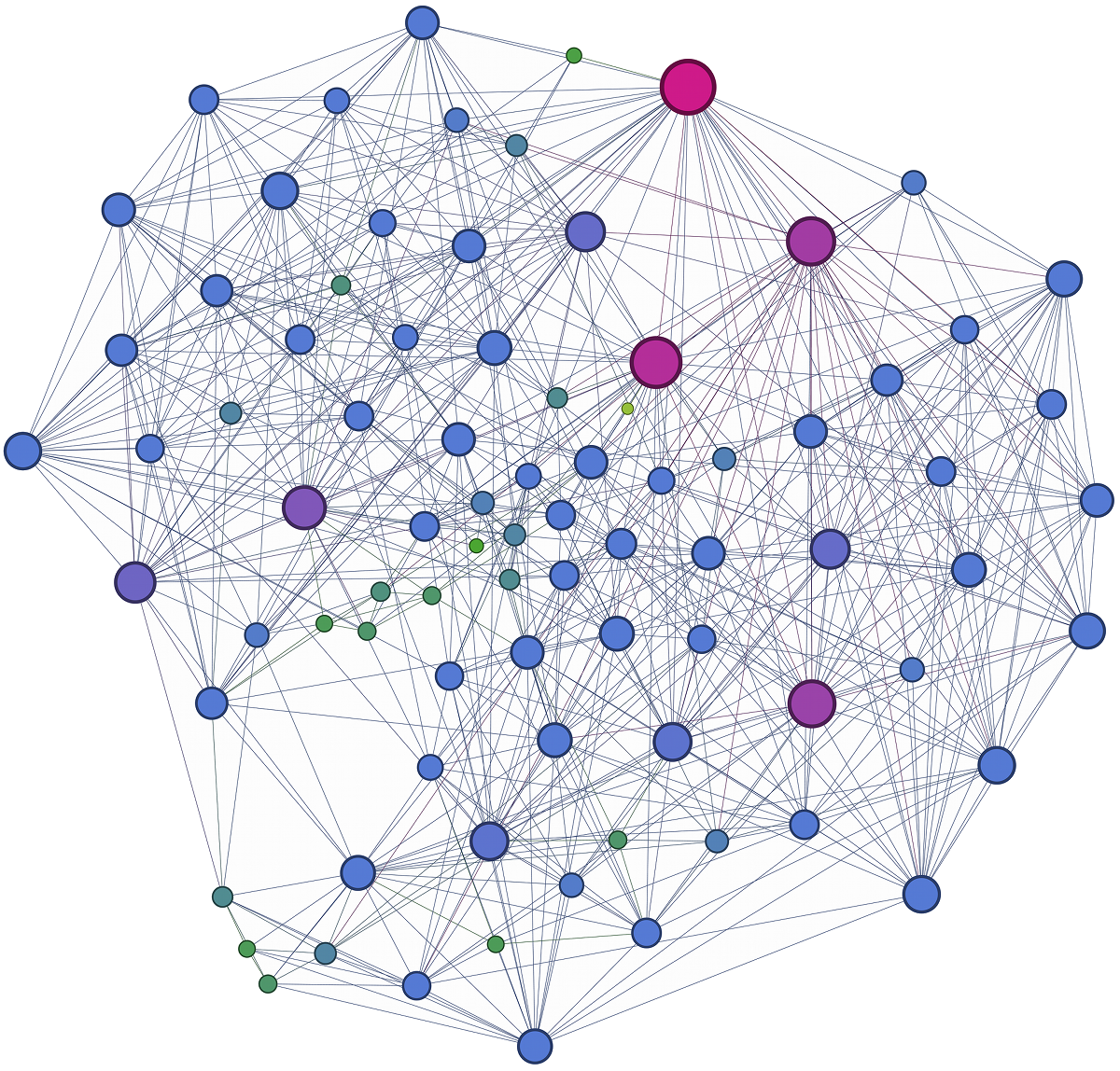}}
  \caption{\textbf{Infectious Contact Network (Dublin) from Section 3}}
\label{fig:tscc-reach-vis}
\end{figure*}

\end{document}